\documentclass[aps,pre,twocolumn,reprint,showpacs,superscriptaddress,preprintnumbers,amsmath,amssymb,floatfix,raggedbottom]{revtex4-1}
\usepackage{amsmath, amsfonts, amssymb, amsthm, euscript, amscd, latexsym,varwidth}
\usepackage[misc,geometry]{ifsym}
\usepackage{tabularx}
\usepackage{booktabs}
\usepackage{natbib}
\usepackage{multibib}
\usepackage[utf8]{inputenc}
 \usepackage[breaklinks=true,colorlinks=true,linkcolor=blue,citecolor=blue,urlcolor=blue,hypertexnames=false,bookmarks=false]{hyperref} 
\usepackage[all]{hypcap} 
\usepackage{overpic}

\usepackage{multirow}
\usepackage{epsfig}
\usepackage{dcolumn}
\usepackage{graphicx}
\usepackage{graphics}
\usepackage{color}
\usepackage[normalem]{ulem}
\usepackage[utf8]{inputenc}
\usepackage[T1]{fontenc}

\usepackage{tikz}
\usetikzlibrary{arrows}

\def\aa#1{ \begin{align*} #1 \end{align*} }
\def\aaa#1{ \begin{align} #1 \end{align} }

\newcommand{\lb}{\label}
\newcommand{\rf}{\eqref}

\renewcommand{\pl}{\partial}

\newcommand{\bi}{\begin{itemize}}
\newcommand{\ei}{\end{itemize}}

\newcommand{\sg}{\sigma}

\newcommand{\om}{\omega}
\newcommand{\mc}{\mathcal}
\newcommand{\Om}{\Omega}

\newcommand{\td}{\tilde}

\newcommand{\x}{\times}

\newcommand{\Te}{\Theta}
\newcommand{\te}{\theta}

\newcommand{\nab}{\nabla}

\def\Rnu{{\mathbb R}}

\long\def\symbolfootnote[#1]#2{\begingroup%
\def\thefootnote{\fnsymbol{footnote}}\footnote[#1]{#2}\endgroup}
\newcommand*\colvec[1]{\begin{pmatrix}#1\end{pmatrix}}

\include{srctex.sty}

\begin{document}

\date{\today}

\title{Backward stochastic differential equation approach to modeling of gene expression}
\author{Evelina Shamarova}
\email{evelina@mat.ufpb.br}
\affiliation{
Departamento de Matem\'atica, Universidade Federal da Para\'iba, 58051-900, Jo\~ao Pessoa, Brazil}
\author{Roman Chertovskih}
\affiliation{
Samara National Research University
Moskovskoe shosse 34, Samara
443086 Russian Federation
}
\author{Alexandre Ramos}
\affiliation{
Escola de Artes, Ci\^encias e Humanidades, 
Universidade de S\~ao Paulo, Av. Arlindo B\'ettio 1000, 
03828-00, S\~ao Paulo, SP, Brazil
}
\author{Paulo Aguiar}
\affiliation{
INEB - Instituto de Engenharia Biom\'edica
i3S - Instituto de Investiga{\c c}\~ao e Inova{\c c}\~ao em Sa\'ude,\\
Rua Alfredo Allen 208, 4200-135 Porto, Portugal
}


\begin{abstract} 
In this article, we introduce a novel backward method to model
stochastic gene expression and protein level dynamics. The protein amount is
regarded as a diffusion process and is described by a backward stochastic
differential equation (BSDE). Unlike many other SDE techniques proposed in the
literature, the BSDE method is backward in time; that is, instead of initial
conditions it requires the specification of endpoint  (``final'') conditions, in
addition to the model parametrization. To validate our approach we employ
Gillespie's stochastic simulation algorithm (SSA) to generate (forward) benchmark
data, according to predefined gene network models. Numerical simulations show that
the BSDE method is able to correctly infer the protein level distributions that
preceded a known final condition, obtained originally from the forward SSA. This
makes the BSDE method a powerful systems biology tool for time reversed
simulations, allowing, for example, the assessment of the biological conditions
(e.g. protein concentrations) that preceded an experimentally measured event of
interest (e.g. mitosis, apoptosis, etc.).  
\end{abstract}

\maketitle

\section{Introduction} 
Gene regulatory networks involving small numbers of
molecules can be intrinsically noisy and subject to large protein concentration
fluctuations \cite{Thattai2001, Ozbudak2002}. This fact substantially limits the
ability to infer the causal relations within gene regulatory networks and the
ability to understand the mechanisms involved in healthy and pathological
conditions. A large interest has been raised in developing tools for gene
regulatory network inference \cite{Karlebach2008,Hecker2009} acknowledging the
noisy/stochastic properties of experimental data \cite{Mettetal2006,
Wilkinson2011,Lillacci2013}, in parallel with studies addressing the
prospective, forward, simulation of stochastic equations describing biochemical
reactions \cite{Gillespie1977}. There is, however, another context which, despite
its relevance as a tool to better understand intracellular dynamics, has
received little attention from a mathematical modeling perspective. That is the
situation where the basic gene regulatory network is known, together with a
present distribution of molecules/proteins, and one wants to infer the previous
molecules distributions that gave rise to the observed data. This is the case,
for example, of a sample of necrotic cells where the concentration distributions
for the relevant molecules can be calculated, and one would like to infer the
previous concentrations that gave rise to the necrotic condition. In this
context, the problem can be addressed with backward stochastic
differential equations.

BSDEs were introduced by Bismut in 1973 \cite{Bismut}, 
and over the last twenty years have been extensively studied 
by many mathematicians (e.g. \cite{Pardoux1}, \cite{Ma}). 

In what follows, we present a method to model gene expression based on backward
stochastic differential equations.  We consider a gene regulatory network,
where the stochastic variables are the amounts of proteins that are expressed from
the genes of the network.  To illustrate our method, we apply it to four simple
gene networks: a positive self-regulating gene, which is the simplest network, 
networks composed by two and five interacting genes, and a bistable two-gene network.
To generate data to test
and validate our approach we use Gillespie's stochastic simulation algorithm,
referred to below as SSA (\cite{Gillespie1977}), for simulation of biochemical reactions.  From the trajectories of multiple simulations, the SSA
provides the distribution of protein amounts at a fixed final time, as well as at
some fixed moments of time prior to the final. For realization of the SSA we used
the  COPASI software \cite{copasi}.  The network models used in the BSDE and the
SSA simulations were taken the same. The BSDE method, which requires the final
distribution as the input data, was applied to perform a simulation backwards in
time. Importantly,  at the end of the backward simulation we arrive at some
deterministic value for the number of proteins which is very close to the SSA
initial condition.  Since in many applications the initial protein amounts are not
known, and are, in fact, the goal of the study, we believe that our approach can be a
useful tool in systems biology.

\section{The BSDE method}

In what follows, we describe the BSDE method to model gene
expression. Specifically, we model the dynamics of protein amounts
expressed by the genes of a gene regulatory network. In our simulation,  the protein synthesis and
degradation occurs on the time interval $[t_0,T]$.  
The input data for the BSDE method is the protein number
distribution at time $T$. 
The amount of proteins is modeled by a continuous $\Rnu^n$-valued diffusion
process $\eta_t = (\eta_1(t), \eta_2(t), \ldots, \eta_n(t))$, where $n$ is the
amount of species, or types of proteins expressed by the genes of the network,
and $\eta_i(t)$ is the amount of the $i$-th type of protein at time $t$.  

In our model, the transcription and
translation are treated effectively as a single process. 
In
other words, we assume that different mRNAs transcribed from the gene are
translated at the same rate. 

\subsection{General description of the method}
\lb{method} 

In the BSDE method,
the evolution of $\eta_t$ is governed by the
following BSDE
\aaa{
\lb{bsde2}
\eta_t = \eta_T - \int_t^T f(\eta_s)\, ds - \int_t^T z_s\, dW_s,
\; t\in [t_0,T].
}
On the right-hand side, $\eta_T$ is the vector of final amounts of proteins whose distribution at time $T$ is known,
the second term is a drift that represents the regulation of the protein production,
and the last term is an unknown noise that makes the solution $\eta_t$ stochastic.
Furthermore, 
$W_s$ is a real-valued Wiener process (also referred to below as a Brownian motion), 
and $f$ is an $\Rnu^n$-valued 
synthesis/degradation rate of the proteins under regulation whose explicit form is discussed in detail in Section \ref{fun}.
Rigorously speaking, the last term in \rf{bsde2} is
an It\^o stochastic integral with respect to the Brownian motion  $W_s$,
where the integrand $z_s$ is an unknown stochastic process.

In order to solve BSDE \rf{bsde2} numerically we represent
$\eta_T$ in the form $h(W_T)$, where $h(x)$ is a continuous function defined for real values $x$ 
and taking values in $\Rnu^n$. This function will be
obtained numerically during the realization of our method. 
The main tool for obtaining a numerical solution to \rf{bsde2}
is the following deterministic final value problem with respect to an unknown
$\Rnu^n$-valued function $\te(t,x)$ defined for $(t,x) \in [t_0,T]\x \Rnu$:
\aaa{
\lb{pde}
\begin{cases}
\pl_t \te(t,x) + \frac12\te_{xx}(t,x) - f(\te(t,x)) = 0,\\
\te(T,x) = h(x), \quad x\in \Rnu.
\end{cases}
}
In the above PDE, the variable $x$ is an abstract variable that is to be substituted by a 
Wiener process to generate the solution to equation \rf{bsde2}, and PDE \rf{pde} itself is a \textit{tool} to obtaining
a solution to BSDE \rf{bsde2}.
Namely, the theory of BSDEs (\cite{Ma}) implies that if $\te(t,x)$ 
is a solution to problem \rf{pde}, then the pair of stochastic processes
\aaa{
\eta_t=\te(t,W_t) \quad \text{and} \quad z_t=\nab \te(t,W_t)
}
is the unique solution to \rf{bsde2} under the constraint that $\eta_t$ is adapted with respect to $W_t$
(see  \cite{Pardoux1}, \cite{Ma} for details).
The forementioned adaptedness means
that for each $t$, $\eta_t$ is a function of $W_t$.
We provide more details about BSDEs in the appendix.


Let us summarize the algorithm of obtaining a numerical solution to BSDE \rf{bsde2}.
 (a) Construct the function $h$ with the property $h(W_T) = \eta_T$; (b) Obtain a numerical solution $\te(t,x)$ to problem \rf{pde};
 (c) Simulate a sufficient number of Brownian motion trajectories and obtain the solution to \rf{bsde2} in the form $\eta_t = \te(t,W_t)$. 

Let us start with (a). 
We obtain the distribution of $\eta_T$ in the form of a histogram $H$.  
The $\Rnu^n$-valued function $h$ is chosen 
so that the distribution of $h(W_T)$
produces a histogram
approximately equal to $H$. The method of finding the function $h$ 
and, therefore, obtaining $\eta_T$ as $h(W_T)$,
is referred to below as
the \textit{final data approximation technique}.

Let $l_i$, $i=1,2, \ldots$, be the bin ends of the given histogram $H$, and $p_i$ be the
bin probabilities. This means that the probability that $\eta_T$ belongs to $[l_i,l_{i+1}]$ is $p_i$.
We search $h$ as a piecewise linear continuous increasing function
of the form
\aaa{
\lb{fh} 
h(x) = \sum_{i=1}^N \chi_{[r_i,r_{i+1})}(x) (k_i x + b_i)
}
where $\chi_{(r_i,r_{i+1}]}(x)$ is the characteristic function the interval $[r_i,r_{i+1})$, i.e.
$\chi_{[r_i,r_{i+1})}(x) = 1$ if $x$ belongs to $[r_i,r_{i+1})$ and it is zero otherwise. 
We aim to choose $k_i$ and $b_i$ so that $h(r_i) = l_i$, i.e.
$h$ maps $[r_i,r_{i+1}]$ onto $[l_i,l_{i+1}]$.
Since $\eta_T$ is in $[r_i,r_{i+1}]$ with probability $p_i$, the forementioned property of $h$ implies that $W_T$ belongs to $[r_i,r_{i+1}]$ also with probability $p_i$. 
Thus, we produce 20000 realizations of the random variable $W_T$.
The endpoint $r_1$ is choosen as the smallest of the realizations of $W_T$. Suppose
we constructed the endpoint $r_{i}$. 
Note that $W_T$ is a Gaussian random variable with mean zero
and variance $\sqrt{T}$. Let $\Phi_T(x)$ be the distribution function of $W_T$. Clearly,
we can uniquely find the point $r_{i+1}$ so that $\Phi_T(r_{i+1}) - \Phi_T(r_i) = p_i$.
Further, we compute $k_i = (l_{i+1} - l_{i})/(r_{i+1} - r_i)$ and choose $b_i$ 
so that $h(x)$ becomes continuous at point $r_i$, i.e. $b_i = r_i(k_{i-1}- k_i) + b_{i-1}$.
Since computing of  $b_1$ requires $b_0$, we set $b_0$ to be  the mean of $\eta_T$.

We remark that continuous function $h$ satisfying $h(W_T) = \eta_T$ may not be unique. However,
the goal of the construction of $h$ is to be able to solve BSDE \rf{bsde2} by means of problem \rf{pde}.
From the theory of BSDEs (\cite{Pardoux1}) it is known that the $\mc F_t$-adapted solution pair $(\eta_t,z_t)$ 
is, in fact, uniquely determined by the final data $\eta_T$. 

Now we describe part (b) of the algorithm which is obtaining a numerical solution to \rf{pde}.
By doing the time change $\td \te(t,x) = \te(T-t,x)$ we transform \rf{pde} to a Cauchy problem with the initial condition
$\td\te(0,x) = h(x)$. Note that, by \rf{fh}, the function $h$ is defined only on a compact interval $[r_1, r_{N+1}]$ which is 
the support for all the realizations of $W_T$. The values of $h$ outside of this interval do not affect the
solution to \rf{bsde2}. Therefore, we can extend $h$ to the whole real line $\Rnu$
so that the extended function is continuous  and its derivative vanishes outside
of a compact interval $[a,b]$ containing $[r_1,r_{N+1}]$. Therefore, in practice, instead of \rf{pde} we solve the following initial-boundary value problem:
\aaa{
\lb{ini-b}
\begin{cases}
\pl_t \td\te(t,x) - \frac12\td\te_{xx}(t,x) + f(\td\te(t,x)) = 0,\\
\td\te(0,x) = h(x), \\
\td \te_x(t,a) = \td \te_x(t,b) = 0.
\end{cases}
}  
Finally, in part (c) we simulate a sufficient number 
of Brownian motion trajectories $W_t$
starting at zero at time $t_0$ 
and obtain the trajectories $\eta_t$ as $\te(t,W_t)$. In our simulation, we took 20000 trajectories of $W_t$.
We note that the noise can be computed as the stochastic integral $\int_t^T\nab \te(s,W_s) dW_s$.
However, as mentioned earlier, in this work we are only interested in the protein amount process  $\eta_t$.

\subsection{BSDE model for multistability}
\label{multi}
Here we extend the model described in \ref{method} to the case when the observed final distribution is bimodal.
For simplicity, we describe the method
for two-gene networks, although our strategy can be naturally extended for networks composed by 
more than two genes. 
The stochastic equation describing the dynamics of proteins synthesis and degradation is still BSDE \rf{bsde2}, however
we decouple it into two BSDEs and solve each BSDE separately.  
Namely, we split the set of random parameters $\Om$ into two disjoint sets $\Om=\Om_A\cup\Om_B$, and
represent the stochastic process $\eta_t = (\eta_1(t),\eta_2(t))$ in the form
\aa{
\eta_t = \colvec{\eta_1(t)\\ \eta_2(t)}  \chi_{\Om_A} + \colvec{\eta_1(t)\\ \eta_2(t)} \chi_{\Om_B}
= \eta^A_t + \eta^B_t,
}
where $\chi_{\Om_C}(\om)$, $C=A,B$, is the characteristic function of the set $\Om_C$  (i.e.
$\chi_{\Om_C}(\om) = 1$ if $\om\in\Om_C$ and $\chi_{\Om_C}(\om) = 0$ otherwise),
 $\eta^A_t = \eta_t  \chi_{\Om_A}$ and $\eta^B_t = \eta_t  \chi_{\Om_B}$. 
 
In fact,  in SSA numerical experiments involving two-gene networks for some rate functions $f(\eta)$ we observed that 
protein amount trajectories $\eta_1(t)$ and $\eta_2(t)$ split into two branches (See Fig. \ref{fig0}). 
\begin{figure}[h]
\begin{overpic}[width=8.5cm]{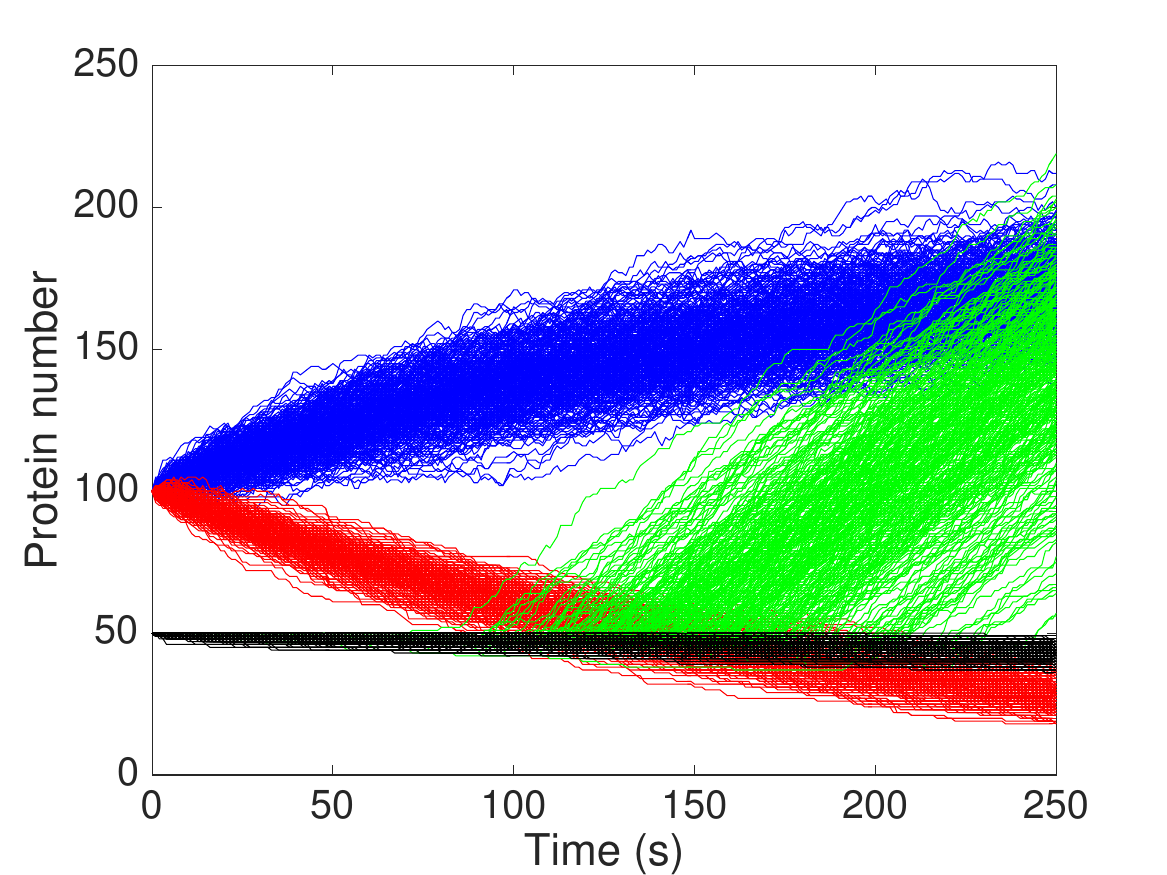}
\put(18,73.6){\fbox{\begin{varwidth}{\textwidth}\textbf{Protein number trajectories:}  \\ \textbf{Multistability in a 2-gene network} \end{varwidth}}}
\end{overpic}
\caption{Observation of bistability in a 2-gene network. The protein number trajectories for the first gene (the initial number 100) split
into the ``blue'' and ``red'' branches. The protein number trajectories for the second gene (the initial number 50) split into the ``black'' and ``green'' branches. 
Each numerical experiment produces either a ``blue'' and a ``black'' trajectory or a ``red'' and a ``green'' trajectory.
The trajectories are obtained by the SSA method.}
\lb{fig0}
\end{figure}
Recall that each experiment (which we regard as a trial and parametrize by a random parameter $\om\in\Om$) produces one trajectory for the first
gene, $\eta_1(t,\om)$, and one trajectory for the second gene, $\eta_2(t,\om)$.
As we repeat the numerical experiment,
the trajectories $\eta_1$ split into the ``red'' and the ``blue'' branches, and
the trajectories $\eta_2$ split into the ``green'' and the ``black'' branches.  
Moreover, the observation shows that whenever a trajectory $\eta_1(t,\om)$ is ``blue'',
the trajectory $\eta_2(t,\om)$ is ``black'', and whenever a trajectory $\eta_1(t,\om)$ is ``red'', the trajectory $\eta_2(t,\om)$ is ``green''.
Based on this observation, we build our BSDE model for bistable gene networks by attributing 
the random parameters from $\Om_A$  to
the blue-black-trajectory experiments, and the random parameters from $\Om_B$
to the red-green-trajectory experiments.

 We decouple BSDE \rf{bsde2} into 
 two independent BSDEs 
 with respect to $\eta^A_t$ and $\eta^B_t$ by
 multiplying the both parts of \rf{bsde2} by the characteristic functions $\chi_{\Om_A}$ and $\chi_{\Om_B}$,
 respectively:
 \aaa{
\lb{bsde-a}
&\eta^A_t = \eta^A_T - \int_t^T  \chi_{\Om_A} f(\eta^A_s)\, ds - \int_t^T  z^A_t\, dW_s,\\
\lb{bsde-b}
&\eta^B_t = \eta^B_T - \int_t^T  \chi_{\Om_B} f(\eta^B_s)\, ds - \int_t^T z^B_t\, dW_s,
}
where $z^A_t = \chi_{\Om_1}z_t$ and $z^B_t = \chi_{\Om_2}z_t$. Above, $\chi_{\Om_A}$ 
and $\chi_{\Om_B}$  are assumed to be
independent
from the Wiener process $W_t$ for any $t\in [t_0,T]$.

Next, we apply our
\textit{final data approximation technique} to obtain the real-valued functions $h_A$ and $h_B$ (taking
values in $\Rnu^2$) so that $h_A(W_T)$ approximates  $\eta^A_T$ and $h_B(W_T)$ approximates $\eta^B_T$.

Further, by employing the BSDE method presented in Section \ref{method},
we obtain the solutions  $(\eta^1_t,z^1_t)$ and $(\eta^2_t,z^2_t)$ to the BSDEs
\aaa{
\lb{bsde-1}
&\eta^1_t =  h_A(B_T) - \int_t^T   f(\eta^1_s)\, ds - \int_t^T  z^1_t\, dW_s,\\
\lb{bsde-2}
&\eta^2_t = h_B(B_T) - \int_t^T   f(\eta^2_s)\, ds - \int_t^T z^2_t\, dW_s.
}
Finally, setting $\eta^A_t = \eta^1_t\chi_{\Om_A}$, $z^A_t = z^1_t \chi_{\Om_A}$, $\eta^B_t = \eta^2_t\chi_{\Om_B}$, $z^B_t = z^2_t \chi_{\Om_B}$,
and multiplying \rf{bsde-1} by $\chi_{\Om_A}$
and  \rf{bsde-2} by $\chi_{\Om_B}$, we obtain that $(\eta^A_t,z^A_t)$ and $(\eta^B_t, z^B_t)$ solve \rf{bsde-a}
and \rf{bsde-b}, respectively.
It remains to remark that summing equations \rf{bsde-a} and \rf{bsde-b} gives original BSDE \rf{bsde2} with $(\eta_t, z_t)$  (defined as
$\eta_t = \eta^A_t + \eta^B_t$ and $z_t = z^A_t + z^B_t$) being its solution. 

\section{Numerical realization}

We employed SSA to produce data for validation of the BSDE method.
Specifically, we performed a number of numerical simulations using the software
COPASI \cite{copasi}, which implements the SSA. The following four cases
were simulated: a self-regulating gene, networks of two and five
interacting genes, and a bistable network of two genes.  For all the networks, the distributions of
protein numbers produced by the two methods,  were compared at two middle time points by analyzing visually the corresponding
histograms plotted jointly, and, where it was possible, by comparing the means and the standard
deviations.  Also, we studied how precise the initial protein numbers for the
SSA were recovered by the BSDE method.

In all simulations, the time is measured in seconds. We used the default options for
numerics of the SSA implemented in COPASI.

At time $T=200$ ($T=250$ for the bistable case), the distribution obtained by the SSA for each type of protein is used to produce a 
 histogram which we take as the input data for our method.

\subsection{Protein production}
\lb{fun}
In equation \rf{bsde2}, the function $f(\eta_t)$ represents phenomenologically the protein synthesis and
degradation. In practice,  the protein synthesis is regulated due to the gene 
interaction with transcription factors. However, for
simplicity, we consider coupled transcription-translation, i.e.  we
neglect the translational regulation, and only take into account the transcriptional regulation. 
The regulatory effect onto gene $i$ is 
represented  by a sigmoidal function multiplied by $\nu_i$.  
Sigmoidal functions have been frequently adopted for  phenomenological modeling of
the transcriptional regulation (see \cite{Mjolsness1991, Alon2007, book:320520, book:561048, Vu-2006, Wang-2007, Kim13, weavmr99}).
Further, we assume that the degradation of each type of protein is of the first order and that
for $i$-th protein it occurs at rate $\rho_i$ \cite{Gibson2000}.
Namely, for two or more genes in the network, the synthesis/degradation rate assumes the form
\aaa{
\lb{drhs}
f_i(\eta) = \nu_i\frac1{1+\exp(-\Theta_i)} - \rho_i \eta_i,
}
where the first term is the rate of proteins synthesis, and the second term is
the proteins degradation rate.  Here $\Te_i =\sum_{j=1}^n A_{ij} \eta_j$,  where
$A_{ij}\eta_j$ represents the net regulatory effect of gene $j$ on gene $i$ with
$A_{ij}$ being the strength of this regulation,
while $\Te_i$ is the total regulatory input to gene $i$.
The $n\x n$ weight matrix
$\{A_{ij}\}$ was the previously introduced in \cite{weavmr99}.
Its element $A_{ij}$  can be
negative, positive, or null, indicating repression, activation or
non-regulation, respectively, of gene $i$ by gene $j$. 
If $\Theta_i$ goes to the negative infinity, 
the synthesis rate tends to zero, and it tends to its maximum value $\nu_i$
for $\Theta_i$ going to the positive infinity.
The exponential term in the denominator appears due to the Arrhenius law with $\Theta_i$ indicating 
the synergestic effect of binding of multiple transcription factors on gene's enhancer (\cite{Reinitz2003}). 

In case of one protein ($n=1$), 
we consider a positive self-regulating gene whose synthesis 
rate is given by a Hill function multiplied by the maximum protein synthesis rate $\nu$ 
\cite{Hill1910,Santillan2008, Bhaskaran2015, Gibson2000} , 
and the degradation rate is a linear function with the rate constant $\rho$:
\aaa{
\lb{rate1}
f(\eta) = \nu\frac{a\eta^2}{1+a\eta^2} - \rho \eta.
}
Here $a$ is a positive constant indicating the strength of the self-regulation.

\subsection{Numerical solution to the PDE}
Problem \rf{ini-b}
is solved numerically using the finite-difference discretization with the
implicit  treatment of the linear terms (the Crank-Nicolson method) and the explicit treatment of
the nonlinear terms. In all computations the time step is taken 
$10^{-4}$, and the uniform spatial grid (including the boundaries) is constituted of
1025 points.  We verified that doubling the spatial and the temporal resolutions
shows no qualitative difference.

\subsection{Self-regulating gene}
We started by simulating the protein level dynamics for a self-regulating gene.
The synthesis/degradation rate  $f(\eta)$, given by \rf{rate1}, was taken
with the parameters $a=1$, $\nu=1$, and $\rho = 0.001$.

The network model for the self-regulating gene is shown on the diagram below 
with $f_s$ and $f_d$ standing   for the synthesis and degradation rates,
respectively.

\begin{tikzpicture}[->,>=stealth',shorten >=1pt,auto,node distance=3cm,
                    thick,main node/.style={circle,draw,minimum size=1.3cm,font=\sffamily\small\bfseries}]

  \node[main node] (1) {Gene};
  \node[main node] (2) [right of=1] {Protein};
  \node[] (3) [right of=2] { $\varnothing$ };

  \path[every node/.style={font=\sffamily\small}]
    (1) edge node {$f_s$} (2)
    (2) edge node {$f_d$} (3)
        edge [bend right] node [swap] { ${ a}$ } (1);
  \end{tikzpicture}      
 
The SSA simulation with 20000 trajectories started at time $t_0 = 0$, and
the values of protein numbers for each trajectory were registered at times
$t=50, 100$, and $200$. 
Next, we represented the SSA data at time $T=200$ in the form of a histogram $H$. 
Using our technique of final data approximation described in Section \ref{method}, 
we found a function $h$,
so that 20000 realizations of the random variable $h(W_T)$ give rise to 
a histogram very close to $H$. 
We took $h(W_T)$ as the final data for BSDE \rf{bsde2}
and applied the BSDE method to simulate 20000 trajectories
backwards in time starting from $T=200$.

\subsection{Networks of interacting genes}
\label{networks}
We tested our method for gene regulatory networks
consisting of two and five genes.
The network models were taken as in the diagram below. 
\vfill \vfill
\begin{tikzpicture}[->,>=stealth',shorten >=1pt,auto,node distance=3cm,
                    thick,main node/.style={circle,draw,minimum size=1cm,font=\sffamily\small\bfseries}]

  \node[main node] (1) {$G_i$};
  \node[main node] (2) [right of=1] {$P_i$};
 \node[main node] (4) [above of=2]{ $P_j$};
  \node[] (3) [right of=2] { $\varnothing$ };
  \node[main node] (5) [above of=1]{$G_j$};
  \node[] (6) [right of=4] { $\varnothing$ };

  \path[every node/.style={font=\sffamily\small}]
    (1) edge node {$f_{s,i}$} (2)
    (2) edge node {$f_{d,i}$} (3) 
        edge [bend left] node [swap] { ${ A_{ii}}$ } (1)
        edge [brown, bend right]  node [swap] { ${A_{ji}}$ } (5)
     (4) edge  [blue, bend left] node   [swap]  { ${A_{ij}}$ } (1)
          edge [bend right] node [swap] { ${ A_{jj}}$ } (5)
    (5) edge node {$f_{s,j}$} (4)
    (4) edge node {$f_{d,j}$} (6); 
  \end{tikzpicture}  

Here gene $i$, denoted by $G_i$, generates proteins of type $i$, which we denote by $P_i$, with the synthesis rate $f_{s,i}$ given
by the first term in \rf{drhs}. Proteins $P_i$ disappear with the
degradation rate $f_{d,i}$ given by the second term in \rf{drhs}.  Proteins $P_i$
have a regulatory effect on gene $j$ (denoted by $G_j$), which is represented by the
regulation coefficient $A_{ji}$.  This holds for any pair $G_i - P_j$.
 In particular, it is assumed that gene $G_i$ generates only protein $P_i$, i.e. gene $G_i$ cannot generate proteins of other types. 
 This means that the
number of genes equals to the number of protein types, i.e. to the dimension of
the random vector $(\eta_1,\ldots, \eta_n)$, where $n$ is either two or five.

For the network of two genes we considered the following values of
parameters: $\nu_1=0.5$, $\nu_2 = 1$, $\rho_1 = 10^{-3}$, $\rho_2 = 5 \cdot 10^{-4}$, 
$A_{11} = 2$, $A_{12} = -1$, $A_{21} = 1$, $A_{22} = 0$.  
For the network of five genes we considered
$\nu  = (0.5, 1, 1, 1, 0.5)$, $\rho =  (10^{-3}, 5 \cdot 10^{-4}, 10^{-3}, 5 \cdot 10^{-4},
10^{-3})$, $A_1 = (2 , -1,  0, 1,  0)$, $A_2 = (1, 0, 0, 0, 2)$, $A_3 = (1, 0, 1, 0, 0)$, 
$A_4 = (0, 0, 1, 1, 1)$, $A_5 = (0, 1, 0, 0, 1)$, where the $i$-th component of $\nu$
is $\nu_i$, the $i$-th component of $\rho$ is $\rho_i$, and $A_i$ denotes the
$i$-th line of the matrix $A$, $i=1,2,3,4,5$. The final time $T$ equals to 200 in both
simulations.

The numerical algorithm  was exactly the same as for the self-regulating gene.
The number of trajectories in both methods was taken 20000. 
Specifically, the SSA simulation started at $t_0 = 0$, and
the values of protein numbers for each trajectory were determined  at times
$t=50, 100$, and $200$. 
The distribution at final time $T=200$ was approximated by $h(W_T)$, and the BSDE method
provided the distributions at $t=50$ and 100, which were compared 
with the distributions of the SSA data. %

\subsection{Bistability}
\lb{bistable}

As we mentioned Section \ref{multi}, in some of the SSA simulations we were able to observe the bistability. It happened, for example,
when we performed the SSA simulation with the following set of parameters: $\nu_1 = \nu_2 = 1$, $\rho_1 = 5\cdot 10^{-3}$,
$\rho_2 = 5\cdot 10^{-4}$, $A_{11} = 1$, $A_{12} = -2$, $A_{21} = -1$, $A_{22} = 1$, and with the initial protein numbers $\eta_1(0) = 100$, $\eta_2(0) = 50$ 
(see Fig. \ref{fig0}).
As before, we considered 20000 trajectories. At the final timepoint $T=250$ we observed a bimodal distribution for both genes. 
As we observe in  Fig. \ref{fig0} the ``blue'' and  the ``red''  branches are
completely separated at $T=250$, while there is a slight overlapping between the ``black'' and the ``green'' branches, 
which was also observed in histograms. 
In our BSDE model for bistability, described in Section \ref{multi},  
we split the set of random parameters $\Om$ into two disjoint subsets $\Om_A$ and $\Om_B$.
Recall that $\om\in\Om$ parametrizes a numerical experiment, and thus, we split the numerical experiments into two groups,
the first parametrized by $\om\in\Om_A$, and the second by $\om\in\Om_B$,
To perform this splitting in practice,
it suffices to separate the  final data based on the observations for the first gene, i.e. to find a threshold completely separating the modes
(e.g. 80 according to Fig. \ref{fig0}).
That is, if at $T=250$ the protein number is bigger than 80 we attribute 
$\om\in\Om_A$ to this experiment, and $\om\in\Om_B$ otherwise.
Thus, we obtain two data sets which are  treated separately by exactly the same procedure that we described in Section \ref{networks},
with the only difference that the timepoints for comparison with the SSA were taken $t=150$ and $200$. 
After we completed the computation for each data set by the BSDE method, we joined the data from two computations at timepoints $t=150$ and $t=200$.

\section{Results}

\paragraph{Self-regulating gene.} 
In Fig.~\ref{Approx-final-data-1}, we show the histogram $H$ for the SSA data and 
its approximation $h(W_T)$ at time $T=200$ which
demonstrates that our final data 
approximation technique is quite precise. 
\begin{figure}[th]
\epsfig{file=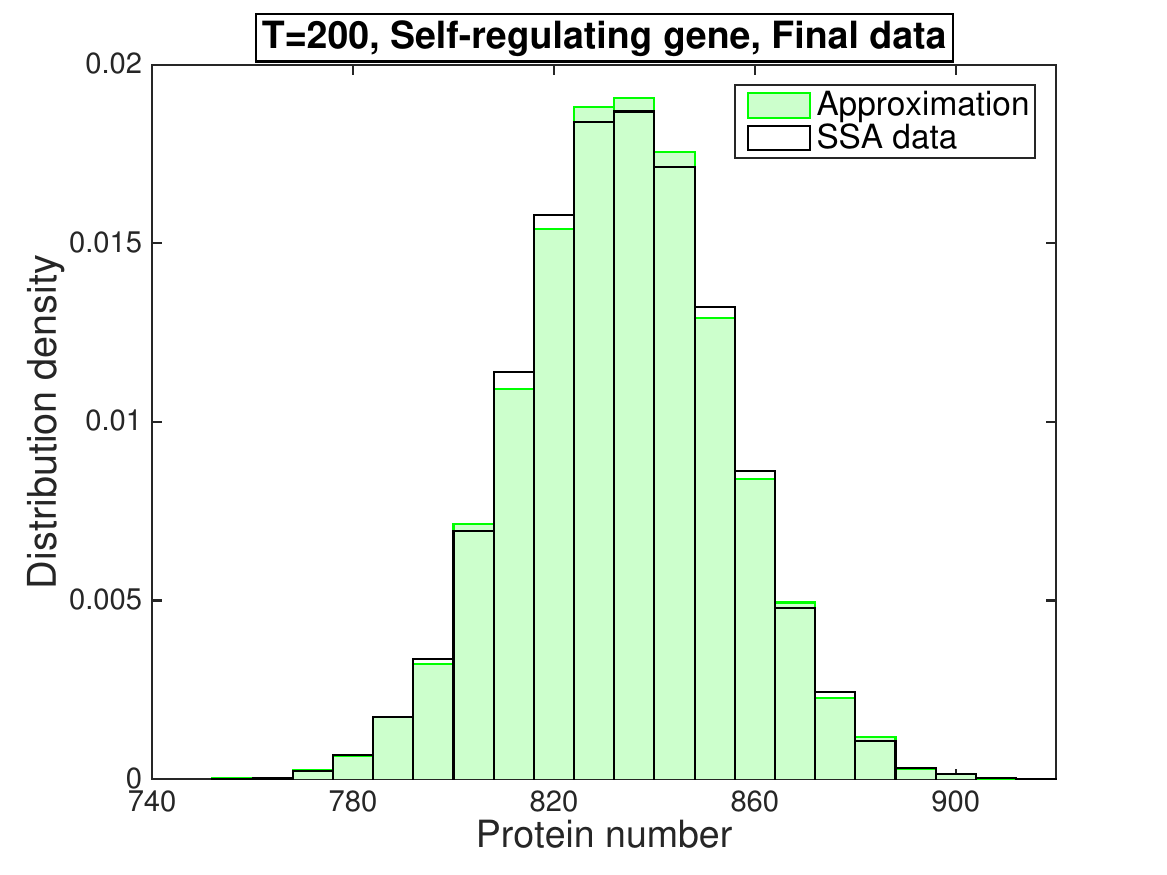,width=8.5cm}
\caption{Histograms for the SSA data and for the approximation $h(W_T)$ at time $T=200$.}
\lb{Approx-final-data-1}
\end{figure}
The distributions of the protein numbers
were determined at $t=50$ and $t=100$,
and the corresponding histograms were plotted jointly with histograms for the SSA data 
 as shown in Fig. \ref{fig1}.
\begin{figure}[h]
\begin{overpic}[width=8.5cm]{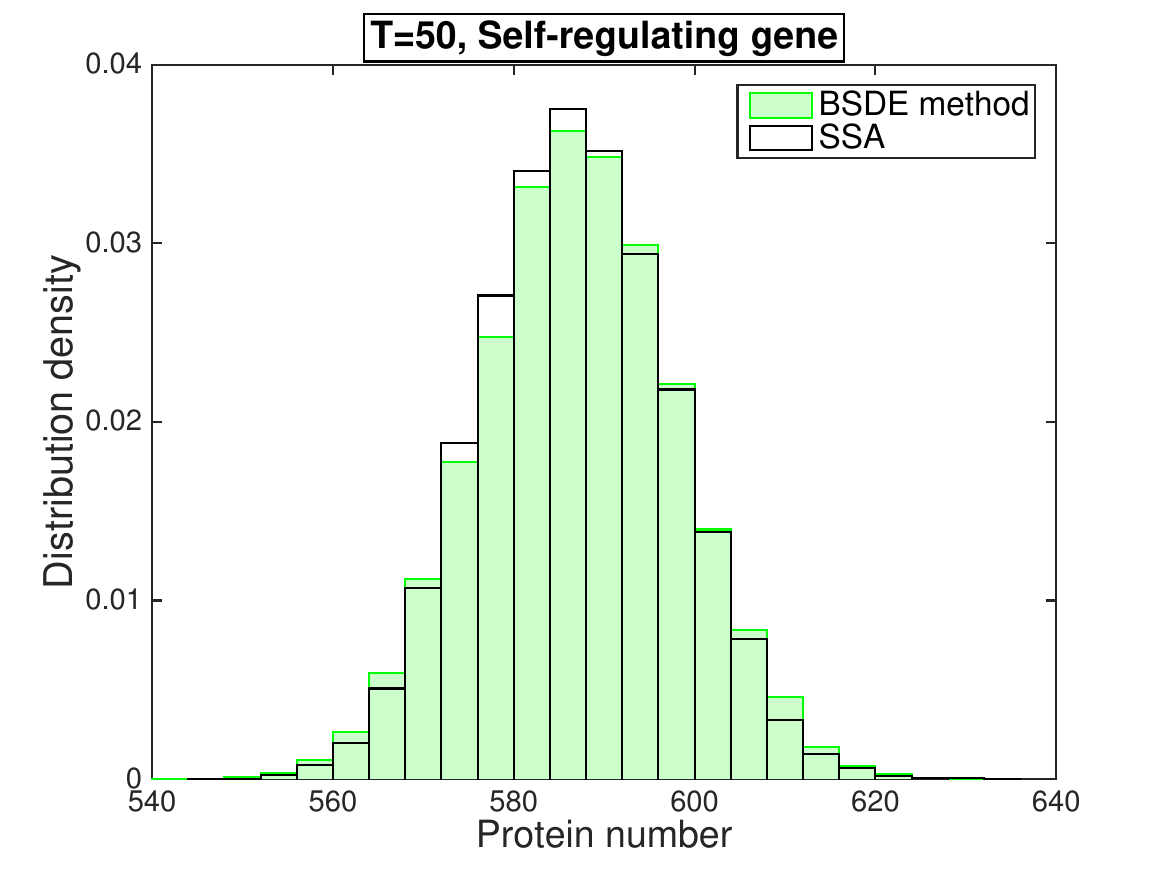}
\put(15,60){(a)}
\end{overpic}
\begin{overpic}[width=8.5cm]{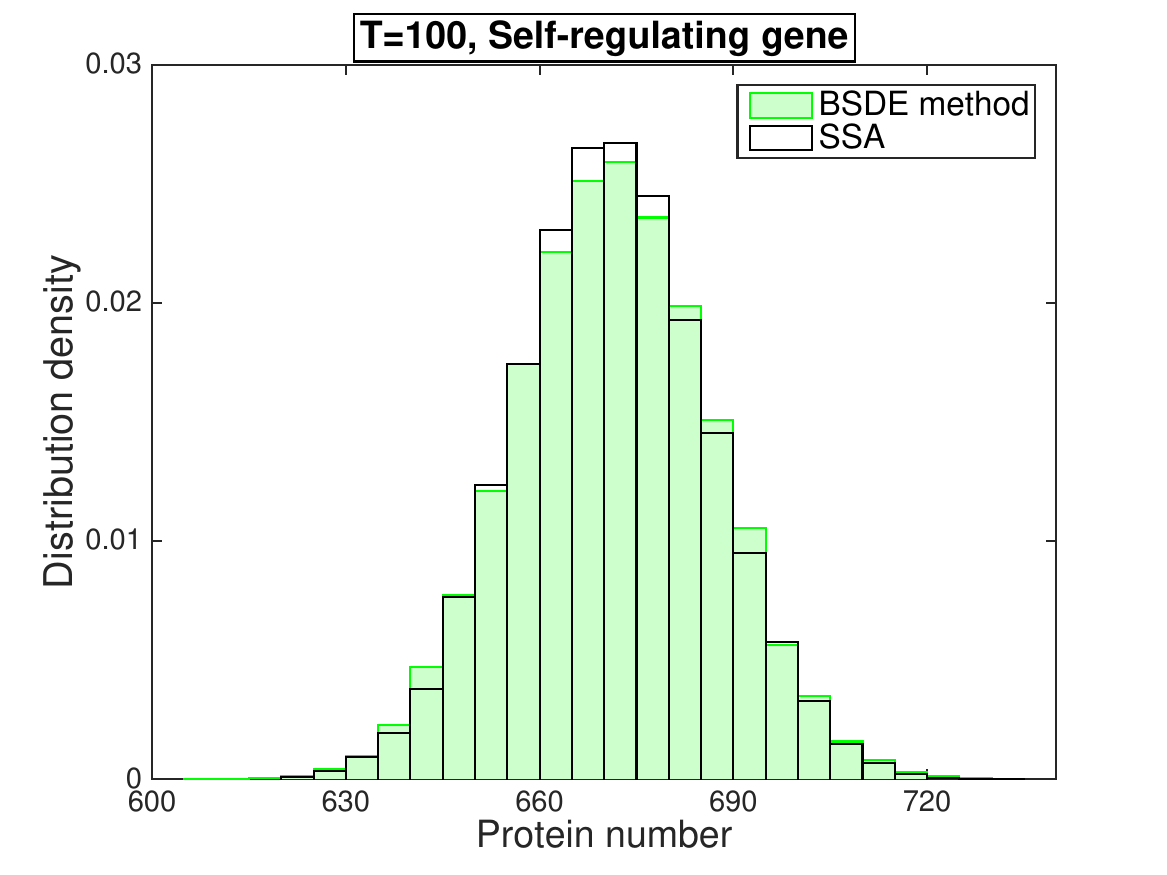}
\put(15,60){(b)}
\end{overpic}
\caption{Distributions of protein numbers for the self-regulating gene at $t=50$ (a) and $t=100$ (b)
for the BSDE method and the SSA.}
\lb{fig1}
\end{figure}

The means $\mu$ and the standard deviations $\sg$ for the data obtained by the  both methods 
are presented in Table \ref{table:tab1}. Although obtained by very different methods, 
the means and the standard deviations are in good agreement.
The percent difference errors were computed as follows:
\begin{gather*}
{\rm Err}\, \mu = |(\mu_{\rm SSA} - \mu_{\rm BSDE})/\mu_{\rm SSA}|,\\
{\rm Err} \,\sg = |(\sg_{\rm SSA} - \sg_{\rm BSDE})/\sg_{\rm SSA}|.
\end{gather*}

\begin{table}[ht]
\centering
\caption{The means $\mu$ and standard deviations $\sg$ for the 
distribution of protein numbers for a 
self-regulating gene computed at $t=$ 0, 50, and 100. 
At $T=200$ we present $\mu$ and $\sg$ obtained by using the final data 
approximation technique in comparison with the SSA data.
The data obtained by the BSDE method are in the second and the third columns, 
and the data obtained by the SSA are in the fourth and the fifth columns. 
The last two columns present the percent difference errors.}
\begin{tabularx}{0.475\textwidth}{X  X  X  X  X  X  X} 
\hline
\hline
& \multicolumn{2}{ c }{BSDE} & \multicolumn{2}{c}{SSA} & \multicolumn{2}{c}{$\%$Errors} \\
\hline
Time & $\mu$ & $\sg$ & $\mu$ & $\sg$ & Err $\mu$ & Err $\sg$ \\
\hline
0 &   500.75 & 0 & 500 & 0 &  0.15$\%$ & --\\
\hline
50 & 587.13 & 10.99 & 586.44 & 10.53 & 0.11$\%$ & 4.35$\%$\\
\hline
100 & 671.37 & 15.30 & 670.78 & 14.78 & 0.08$\%$ & 3.51$\%$\\
\hline
200 & 833.80 & 20.46 & 833.27 & 20.52 & 0.06$\%$ &0.31$\%$ \\
\hline
\hline
\end{tabularx}
\label{table:tab1} 
\end{table}

Some trajectories of the BSDE solution $\eta_t$, representing 
the evolution of the number of proteins generated by a self-regulating gene, are
shown in Fig.~\ref{fig2}.  This is an
illustrative example of what the output of the BSDE method looks like, and how the
trajectories of $\eta_t$ return to the same point which is close to $500$. This is a good approximation
of the protein number that we used as the starting point for the SSA, and therefore,
the prediction of this number by the BSDE method is very precise.

\begin{figure}[th]
\epsfig{file=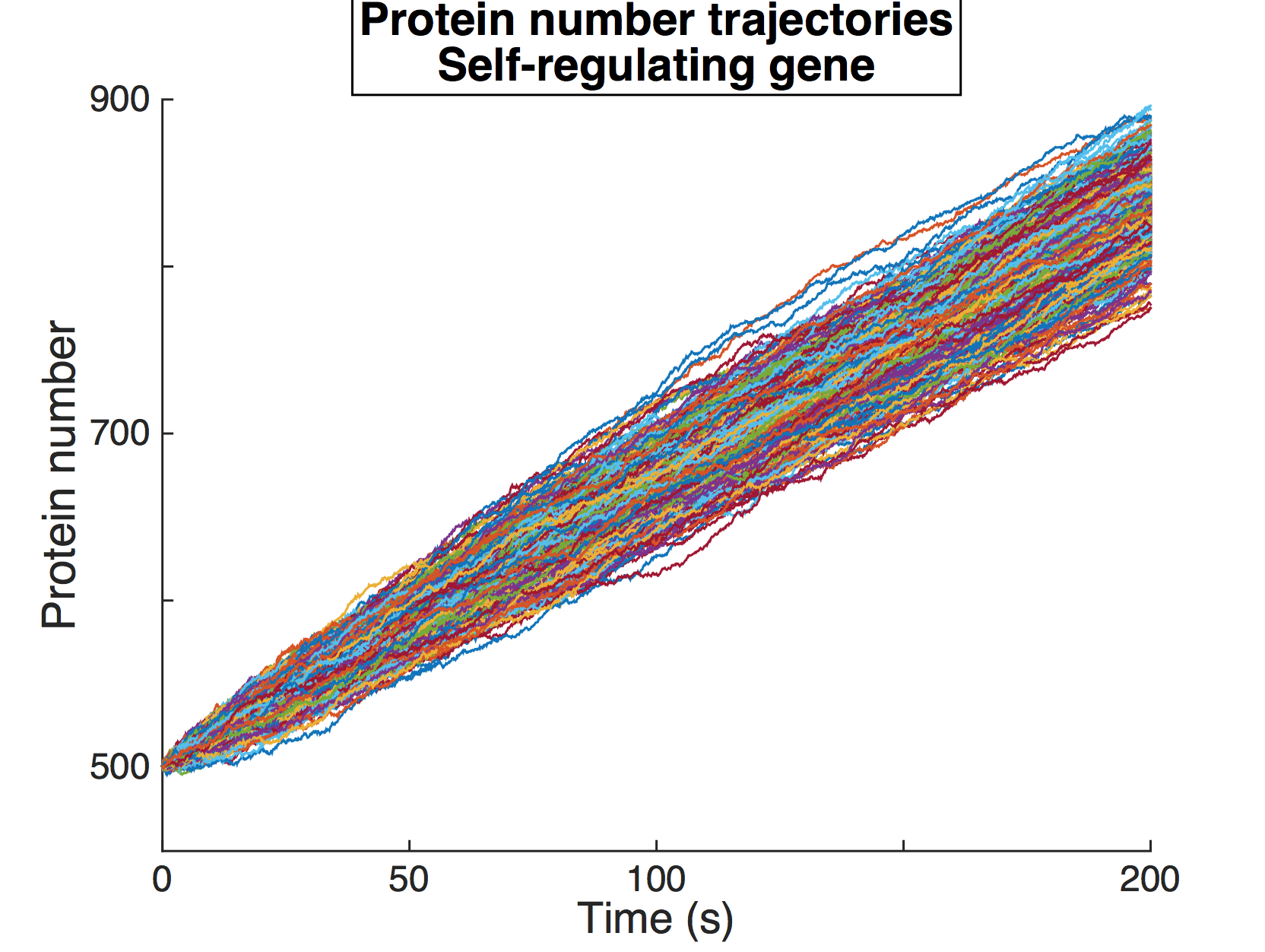,width=8.5cm}
\caption{Trajectories of the stochastic process $\eta_t$, describing the protein 
number for the self-regulating gene, obtained by the BSDE method.}
\lb{fig2}
\end{figure}

\paragraph{Networks of interacting genes.} 
In Figures \ref{fig4} and \ref{fig5} we show the 
distributions at $t=50$ and 100 for some genes of the networks
of two and five genes, respectively.

\begin{figure}[h]
\begin{overpic}[width=8.5cm]{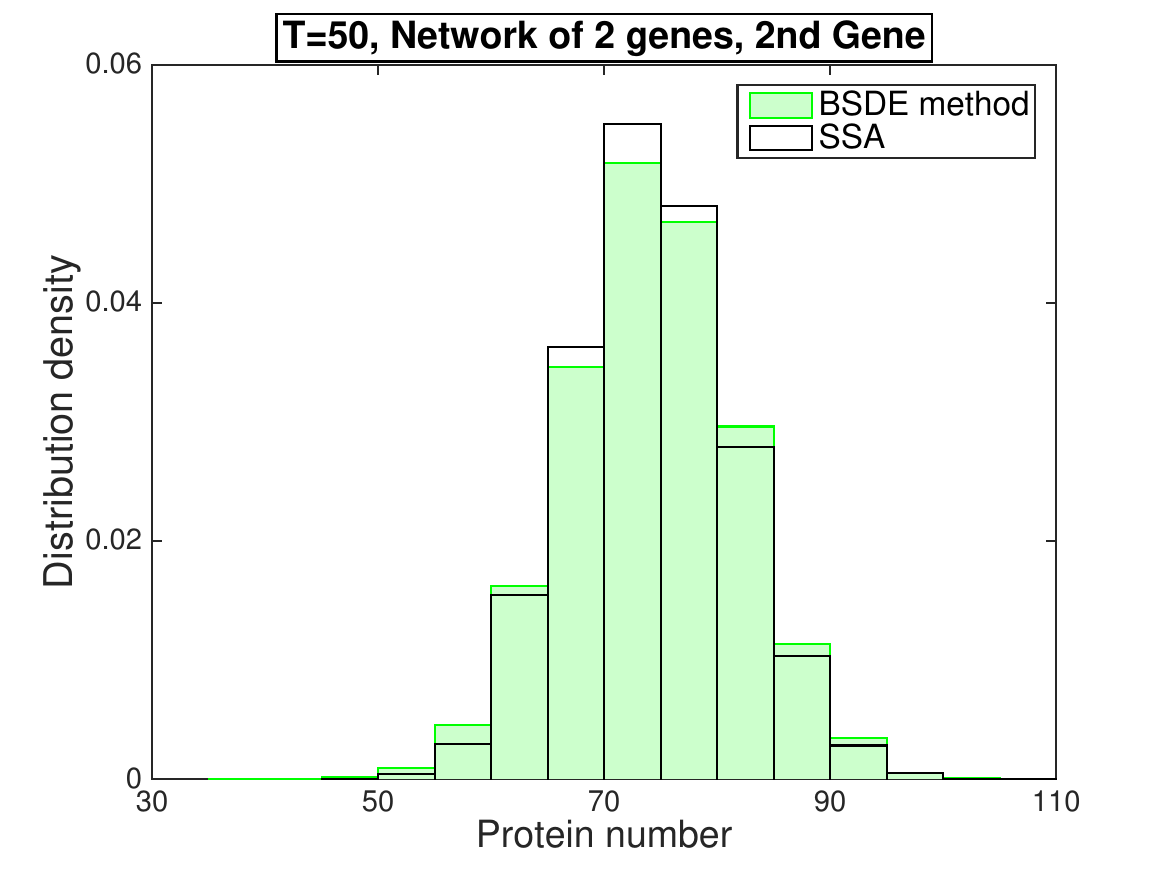}
\put(15,60){(a)}
\end{overpic}
\begin{overpic}[width=8.5cm]{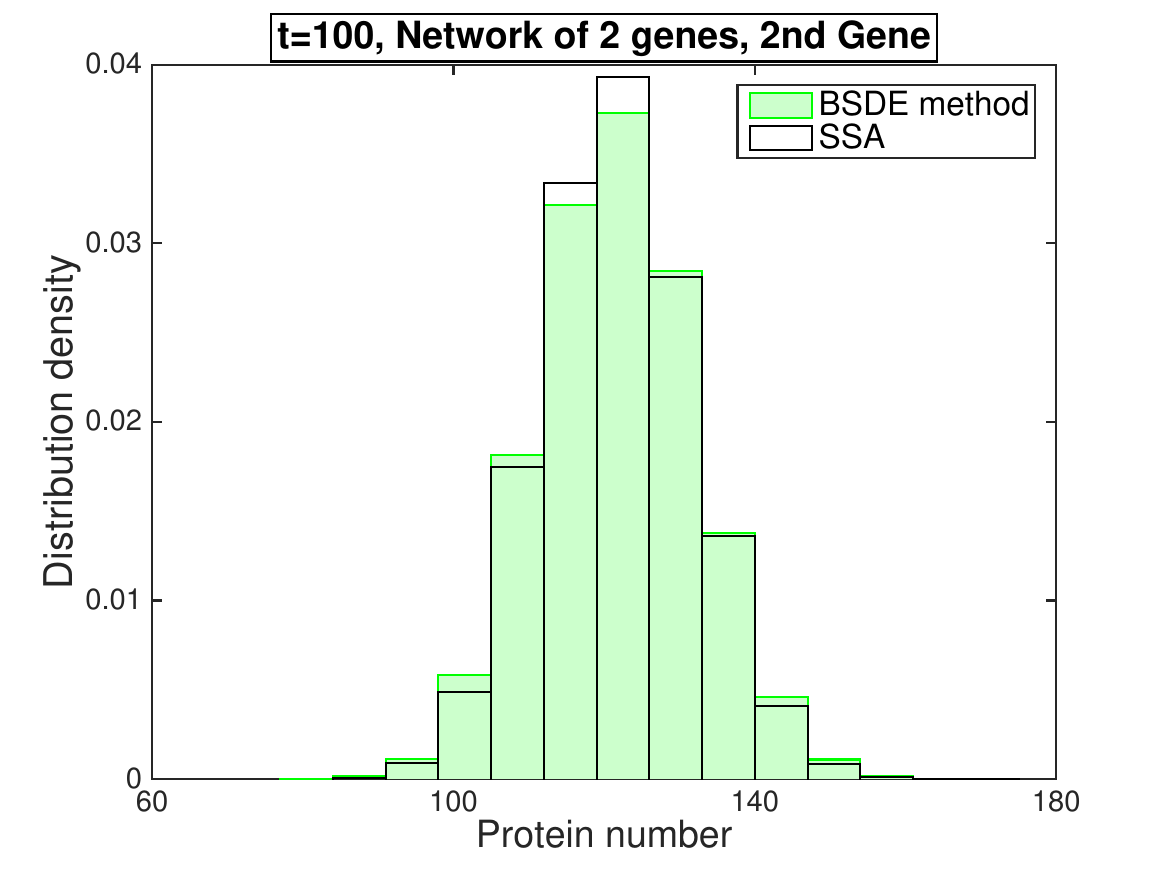}
\put(15,60){(b)}
\end{overpic}
\caption{Distributions of protein numbers at $t=50$ (a) and $t=100$ (b) 
for the 2nd gene of the network of two genes. The distributions are
obtained by the BSDE method and the SSA.}
\lb{fig4}
\end{figure}

\begin{figure}[h]
\begin{overpic}[width=8.5cm]{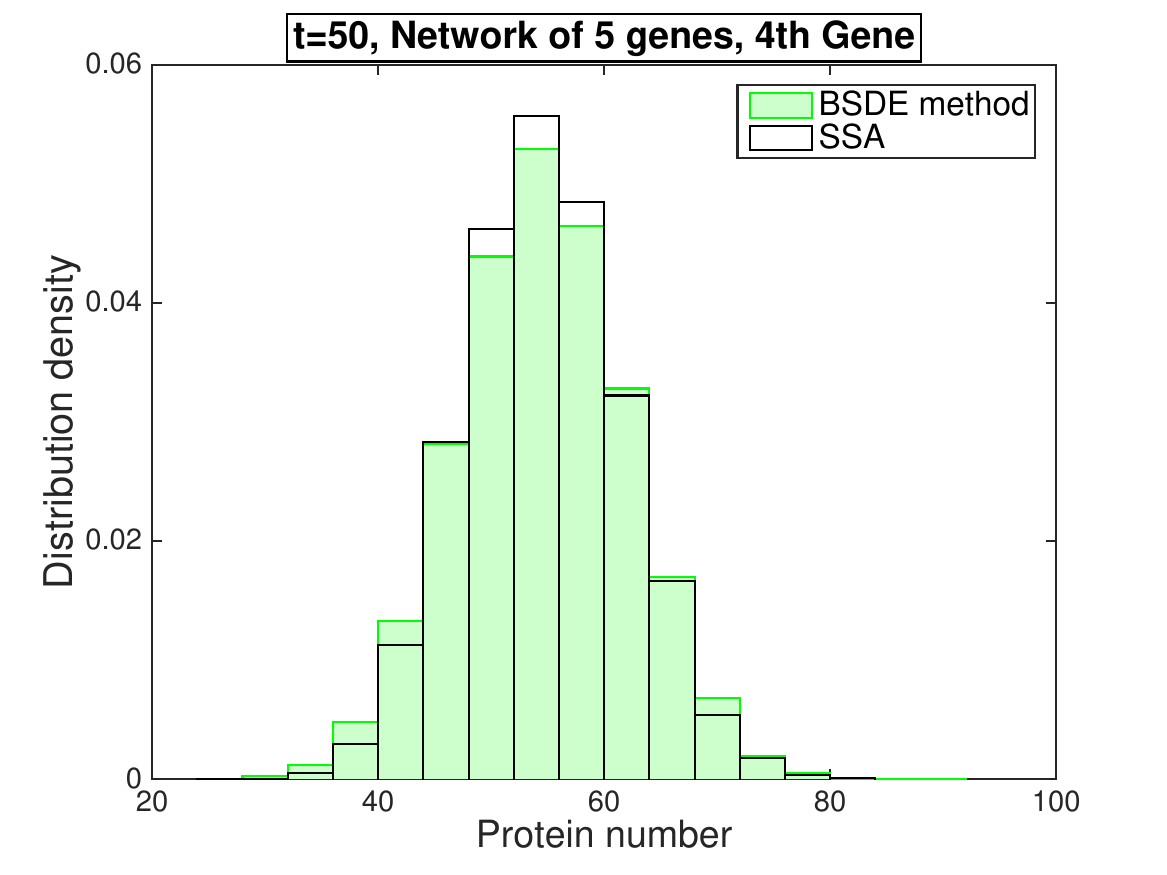}
\put(15,60){(a)}
\end{overpic}
\begin{overpic}[width=8.5cm]{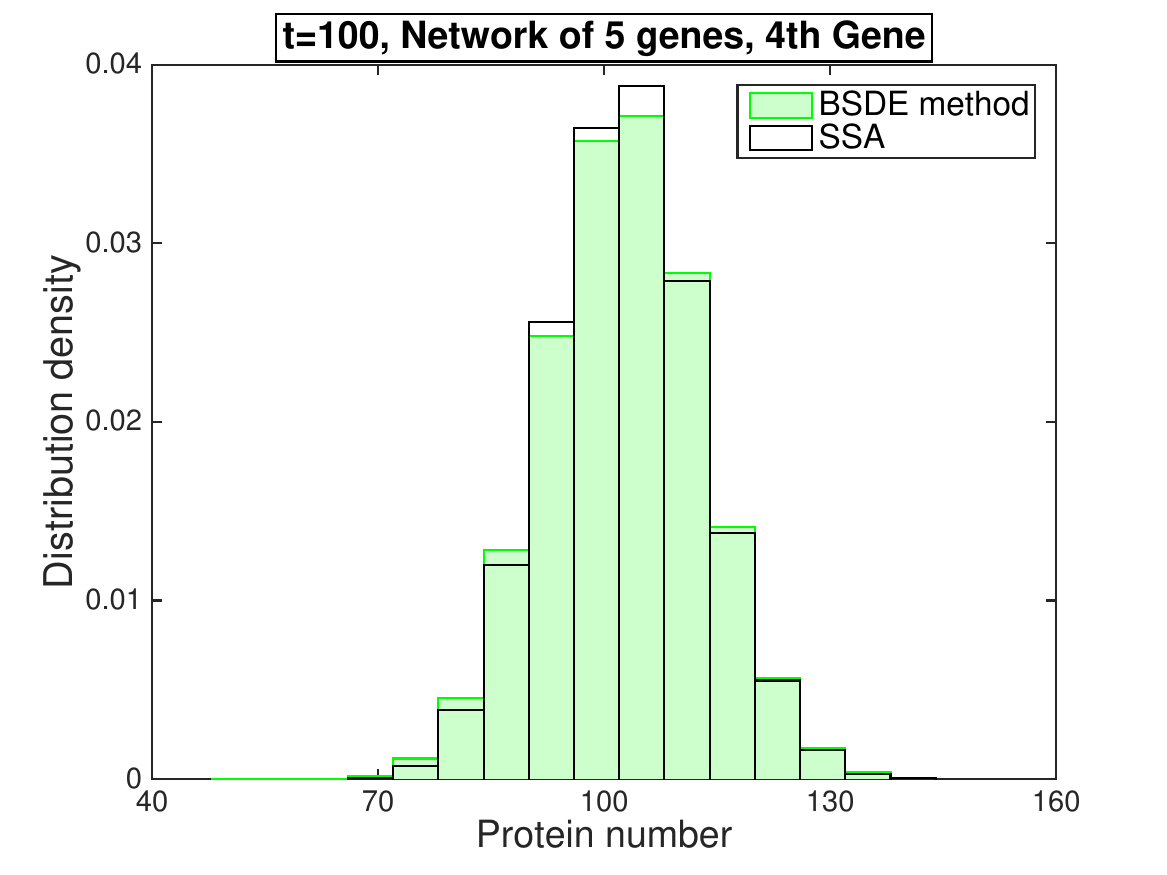}
\put(15,60){(b)}
\end{overpic}
\caption{Distributions of protein numbers at $t=50$ (a) and $t=100$ (b) 
for the 4th gene of the network of five genes. The distributions are
obtained by the BSDE method and the SSA.}
\lb{fig5}
\end{figure}

Also, we compare the means and the standard deviations at $t=50$ and 100
for the data obtained by the both methods. At final time $T=200$ we compare the means
and the standard deviations
obtained by the SSA and by our technique of final data approximation.
The results are presented in Tables \ref{table:tab2} and \ref{table:tab3}.

\begin{table}[ht]
\centering
\caption{The first four columns contain the means $\mu$ and the standard deviations $\sg$ 
for the protein numbers of the network of 2 genes
at $t=0$, $50$, and 100 obtained by the BSDE method and the SSA. 
At $T=200$ we present $\mu$ and $\sg$ obtained by using the final data 
approximation technique in comparison with the SSA data.
The last two columns conintain the percent difference errors.} 
\begin{tabularx}{0.475\textwidth}{X  X  X  X  X  X  X} 
\multicolumn{7}{c}{Network of 2 genes} \\
\hline
\hline
& \multicolumn{2}{c}{BSDE} & \multicolumn{2}{c}{SSA} 
&\multicolumn{2}{c}{$\%$Errors} \\
\hline
$t$ & $\mu$ & $\sg$ & $\mu$ & $\sg$ & Err $\mu$ & Err $\sg$\\
\hline
 \multicolumn{7}{c}{1st gene}\\
\hline
0 &    50.22 & 0 & 50.00 & 0 & 0.44$\%$ & --  \\
\hline
50 &  72.24 &  6.24 & 71.88 & 5.14 & 0.50$\%$ & 21.58$\%$\\
\hline
100 & 92.99 & 8.39 & 92.75 & 7.21 & 0.25$\%$ &$16.41\%$ \\
\hline
200 &  131.85 &  10.81 &  131.23 & 10.94 &  0.47$\%$ & 1.16$\%$  \\
\hline
 \multicolumn{7}{c}{2nd gene}\\
 \hline
0 &    25.43 & 0 & 25.00 & 0 & 1.74$\%$ & --  \\
\hline
50 &  74.29 &  7.54 & 73.79 & 7.10 & 0.67$\%$ & 6.22$\%$\\
\hline
100 & 121.69 & 10.39 & 121.28 & 9.90 & 0.33$\%$ &$4.93\%$ \\
\hline
200 &  213.48 &  13.87 &  212.84 & 13.94 &  0.30$\%$ & 0.58$\%$  \\
\hline
\hline
\end{tabularx}
\label{table:tab2}
\end{table}
\begin{table}[ht]
\centering
\caption{The data representation is the same as in Table~\ref{table:tab2}.}
\begin{tabularx}{0.475\textwidth}{X  X  X  X  X  X  X} 
\multicolumn{7}{c}{Network of 5 genes} \\
\hline
\hline
& \multicolumn{2}{c}{BSDE} & \multicolumn{2}{c}{SSA} 
&\multicolumn{2}{c}{$\%$Errors} \\
\hline
$t$ & $\mu$ & $\sg$ & $\mu$ & $\sg$ & Err $\mu$ & Err $\sg$\\
\hline
 \multicolumn{7}{c}{1st gene}\\
\hline
0 &    50.80 & 0 & 50 & 0 & 1.61$\%$ & --  \\
\hline
50 &  72.70 &  5.75 & 71.90 & 5.19 & 1.11$\%$ & 10.78$\%$\\
\hline
100 & 93.54 & 7.72 & 92.86 & 7.19 & 0.73$\%$ &$7.37\%$ \\
\hline
200 &  132.23 &  9.89 &  131.73 & 9.92 &  0.38$\%$ & 0.32$\%$  \\
\hline
 \multicolumn{7}{c}{2nd gene}\\
 \hline
0 &    25.51 & 0 & 25 & 0 & 2.04$\%$ & --  \\
\hline
50 &  74.25 &  7.52 & 73.78 & 7.14 & 0.63$\%$ & 5.31$\%$\\
\hline
100 & 121.79 & 10.35 & 121.39 & 9.93 & 0.33$\%$ &$4.21\%$ \\
\hline
200 &  213.42 &  13.94 &  212.92 & 13.98 &  0.23$\%$ & 0.28$\%$  \\
\hline
 \multicolumn{7}{c}{3rd gene}\\
 \hline
0 &    10.58 & 0 & 10 & 0 & 5.83$\%$ & --  \\
\hline
50 &  58.82 &  7.77 & 58.31 & 6.98 & 0.89$\%$ & 11.27$\%$\\
\hline
100 & 104.72 & 10.43 & 104.16 & 9.86 & 0.53$\%$ &$5.73\%$ \\
\hline
200 &  189.94 &  13.36 &  189.43 & 13.39 &  0.27$\%$ & 0.28$\%$  \\
\hline
 \multicolumn{7}{c}{4th gene}\\
 \hline
0 &    5.34 & 0 & 5 & 0 & 6.89$\%$ & --  \\
\hline
50 &  54.58 &  7.51 & 54.21 & 7.01 & 0.68$\%$ & 7.04$\%$\\
\hline
100 & 102.61 & 10.33 & 102.27 & 9.95 & 0.33$\%$ &$3.81\%$ \\
\hline
200 &  195.17 &  13.92 &  194.67 & 13.95 &  0.26$\%$ & 0.25$\%$  \\
\hline
 \multicolumn{7}{c}{5th gene}\\
 \hline
0 &    50.66 & 0 & 50 & 0 & 1.32$\%$ & --  \\
\hline
50 &  72.57 &  5.72 & 71.99 & 5.19 & 0.80$\%$ & 10.24$\%$\\
\hline
100 & 93.41 & 7.69 & 92.89 & 7.21 & 0.55$\%$ &$6.58\%$ \\
\hline
200 &  132.12 &  9.84 &  131.61 & 9.87 &  0.38$\%$ & 0.31$\%$  \\
\hline
\hline
\end{tabularx}
\label{table:tab3}
\end{table}

\paragraph{Bistable network of two genes.} 
At timepoints $t=150$ and $t=200$ we
compare the distributions with the SSA in the form of histograms
(see Fig. \ref{fig5}). We observe a good agreement. 
 \begin{figure}[h]
\begin{overpic}[width=8.5cm]{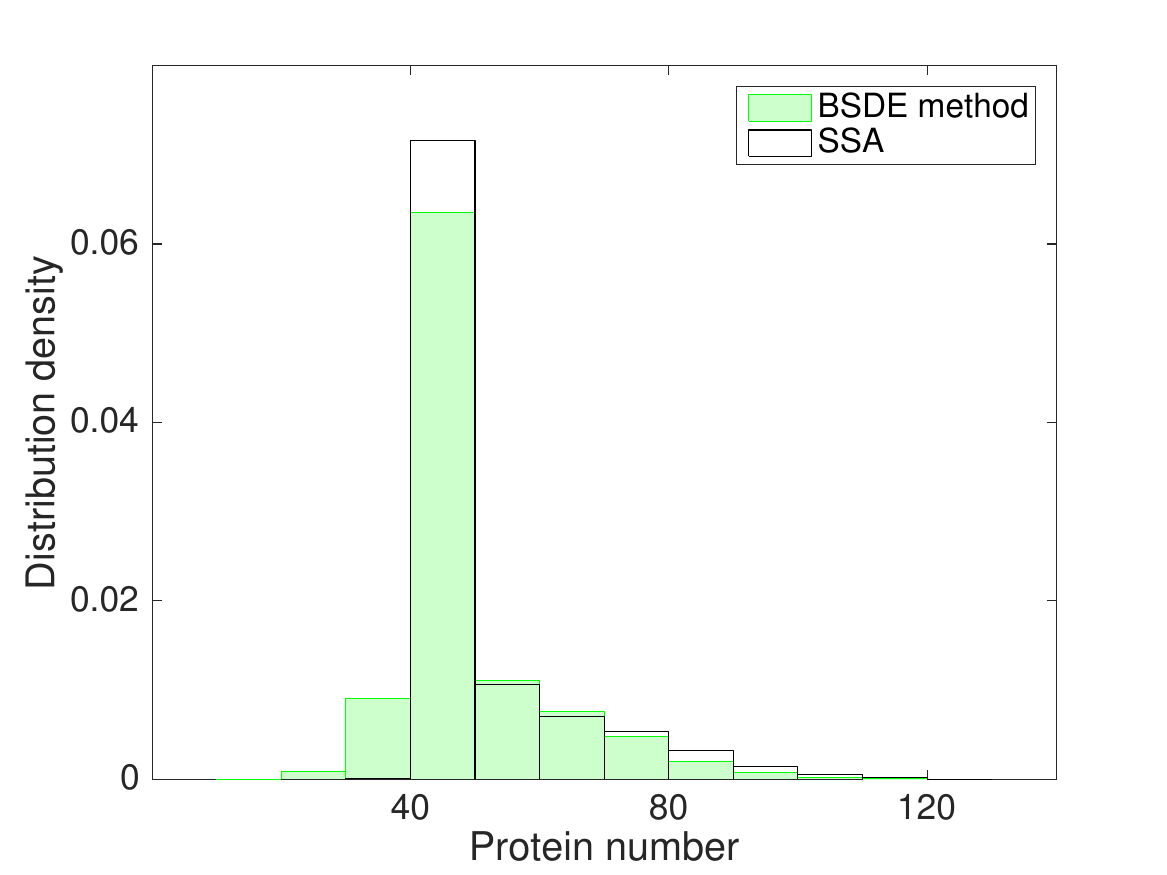}
\put(15,60){(a)}
\put(20,71.5){\fbox{\textbf{t=150, Bistability, 2nd gene}}}
\end{overpic}

\vspace{2mm}

\begin{overpic}[width=8.5cm]{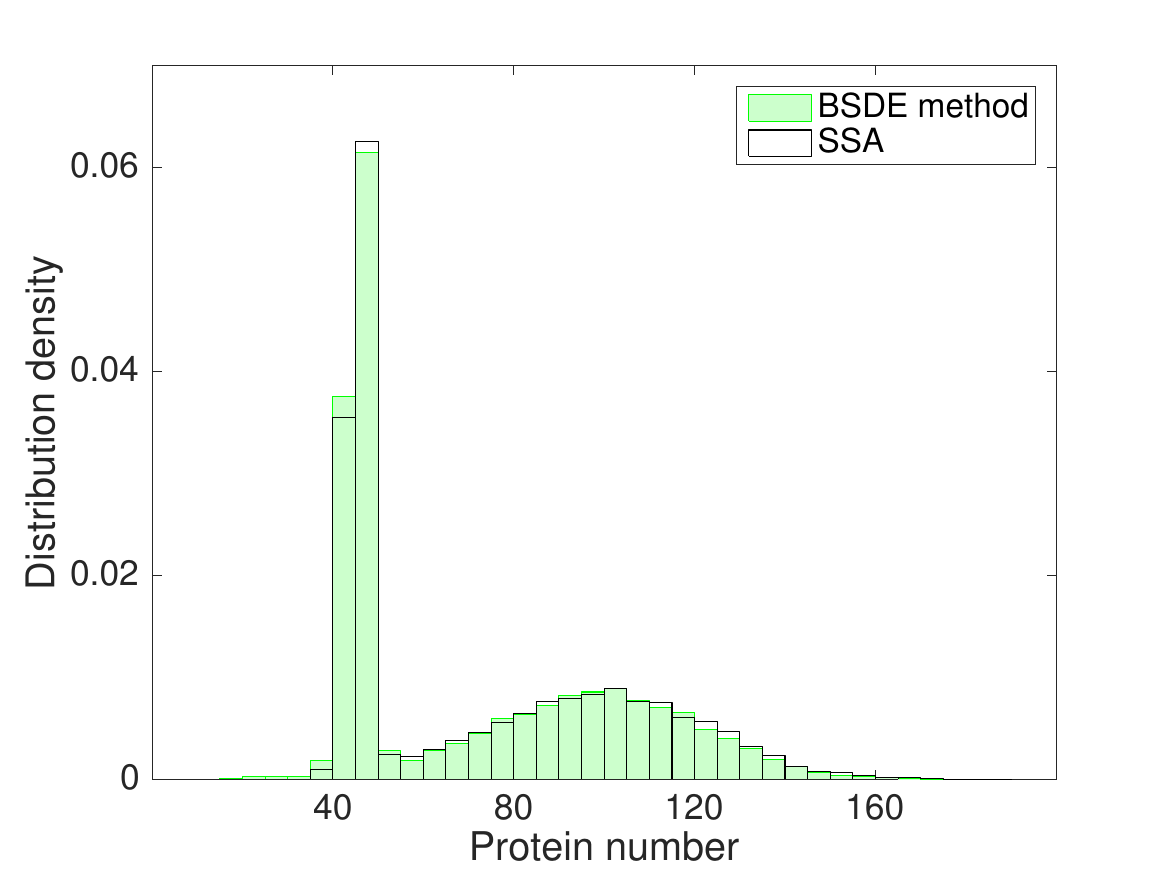}
\put(15,60){(b)}
\put(20,71.5){\fbox{\textbf{t=200, Bistability, 2nd gene}}}
\end{overpic}
\caption{Distributions of protein numbers at $t=150$ (a) and $t=200$ (b) 
for the 2nd gene of the network of 2 genes for the bimodal distribution. 
The distributions are
obtained by the BSDE method and the SSA.}
\lb{fig5}
\end{figure}
Furthermore, we compare the values for initial protein numbers predicted by the FBSDE method with the actual initial protein numbers
used in the SSA simulation. The BSDE simulation of the ``blue'' branch (Fig. \ref{fig0}) provides the initial number
103.83, while the BSDE simulating of the ``red'' branch provides the initial number 100.32 which are close to the initial number used in the SSA simution
(which is 100). Similar results are obtained for the second gene. The results are presented in Table \ref{table:tab5}.
\begin{table}[ht]
\centering
\caption{
The first line contains the data for 
initial protein numbers for the first gene. The value $\mu_A$ 
is obtained by a BSDE simulation of the ``blue'' branch, while 
$\mu_B$ is obtained by a BSDE simulation of the ``red'' branch.
The second line contains the data for 
initial protein numbers for the second gene gene, 
the representation of the data  is similar. The last two columns contain the percent difference errors.} 
\begin{tabularx}{0.475\textwidth}{X  X  X  X  X  X  X} 
\multicolumn{5}{c}{Bistability in 2-genes networks} \\
\hline
\hline
& \multicolumn{2}{c}{\hspace{-1cm}BSDE} &  {SSA} 
&  \multicolumn{2}{c}{$\%$Errors} \\
\hline
Gene & $\mu_A$ & $\mu_B$  & $\mu$ & Err $\mu_A$ & Err $\mu_B$ \\
\hline
1st &    103.83 & 100.32 & 100  & 3.83$\%$  & 0.32$\%$ \\
\hline
2nd &  50.36 &  50.78 & 50 &  0.71$\%$ &  1.55$\%$ \\
\hline
\hline
\end{tabularx}
\label{table:tab5}
\end{table}

\paragraph{Prediction of the initial value.}
As we mentioned before, the BSDE method can be used to approximate the
initial number of proteins. Since the solution to \rf{bsde2} can be represented as
$\eta_t = \te(t,W_t)$, where $\te$ is the solution to final value problem
\rf{pde}, then, as it is implied by the BSDE method, the initial protein number
$\eta_0$ is deterministic and equals to $\te(0,0)$. Tables \ref{table:tab1}--\ref{table:tab5},
show that the BSDE method provides a good
approximation for the initial number of proteins used as an
initial condition in the SSA.
The percent difference error is the biggest, $6,89 \%$, when the initial number
of proteins is $5$ (see Table \ref{table:tab3}), which is the smallest
considered in our simulations. The percent difference error decreases when the
initial protein number increases, and it equals to $0,15 \%$ when we deal with
large initial protein numbers as in the case of the self-regulating gene (see
Table \ref{table:tab1}).

\section{Discussion} 
In this article we presented the BSDE method to model simple
gene expression networks. As a backward method, it relies on the specification of
a gene network model parametrization and on endpoint conditions (as opposed to
initial conditions). It can therefore be applied when we know, or can measure, the
distribution of proteins at a given time, and we want to determine the
distributions at previous time points.  In the BSDE method validation simulations, a
good agreement was found between control and inferred protein level distributions,
in terms of mean values and, in most cases, standard deviations.  The BSDE method
is therefore a powerful tool for time reversed simulations in gene networks /
systems biology, where frequently an endpoint of interest is easily identifiable
(and measured) and the aim is in assessing the prior (causal) conditions.
Another advantage of our method is that it
allows to determine, and even to simulate if necessary, the trajectory of the
noise process. To our knowledge, the noise process is usually unknown and cannot
be determined by any forward method. Obtaining the noise is the subject of our future work.

\paragraph{Determining the final condition.}

The final condition for \rf{bsde2} is required to have the form $h(W_T),$ where
$T$ is the fixed final time. In Section \ref{method}, we described the construction of a piecewise linear function $h$ so that
$h(W_T)$ approximates a given final distribution provided by the SSA
simulation. In practice, to obtain a distribution of protein amounts at time $T$,
a large population of genetically identical cells is usually considered.

\paragraph{Diffusion process approximation.}
We note that the stochastic process describing the
protein number is an integer-valued pure-jump process
which may change its values by $\pm 1$
at time, while the solution to \rf{bsde2} is a continuous process.
However, assuming that the number of proteins of each type is
sufficiently larger than 1, and the waiting times until the next synthesis or
degradation are much smaller than the length of the interval $[t_0,T]$,
we can model the synthesis and degradation of proteins employing continuous
diffusion processes, i.e. by BSDEs with Brownian drivers as \rf{bsde2}.
A diffusion process approximation  for the dynamics of amounts of molecules was undertaken, 
for example, in \cite{Gillespie2000, SDEmodel2005, Ray2008}.

\paragraph{Choice of rate functions.}

We would like to emphasize that the choice of rate functions of form \rf{drhs} is not important for the BSDE method to work.
Although in our simulations we (as well as many other authors 
 \cite{Mjolsness1991, Alon2007, book:320520, book:561048, Vu-2006, Wang-2007, Kim13, weavmr99}) used
  rate functions of form \rf{drhs}, the BSDE method works with any continuous function.

\section{Appendix}
\subsection*{BSDE versus SDE}
\lb{bsde-dis}
%
One may think that BSDE \rf{bsde2} is equivalent to a usual (forward) SDE,
since, similar to ODEs, knowing the final condition instead of the initial
should lead to an equivalent problem.
However, this is not the case if we require the solution to be
adapted with respect to a Brownian motion (i.e. represented as a function of a Brownian motion).
The requirement 
for the pair $(\eta_t,z_t)$ to be adapted implies
that (under some additional analytical assumptions) BSDE \rf{bsde2} 
has a unique solution pair $(\eta_t,z_t)$ \cite{Pardoux1}. Therefore, 
\rf{bsde2} is a different object than the traditional (forward) SDE. 
One may not be convinced why we should require from the solution $\eta_t$ to
be adapted. Gillespie \cite{Gillespie2000} proposed
to model the dynamics of amounts of molecules changing during a chemical reaction by
a forward SDE known as the Chemical Langevin Equation
\aa{
\eta_t = \zeta + \int_{t_0}^t f(\eta_s)\, ds + \int_{t_0}^t z_s\, dW_s,
}
where $\zeta$ is the initial condition at time $t_0$. However, if $\eta_t$ solves
this equation, the theory of SDEs implies that this solution is adapted. The process $z_t$ also must be 
adapted to ensure the existence of the stochastic integral.
Therefore, the requirement
for the solution pair $(\eta_t,z_t)$ to BSDE \rf{bsde2} to be adapted is a natural
consequence of the Langevin dynamics.
\textit{In this article, we propose to use a BSDE for modeling simple gene expression networks
due to its property to have a pair of stochastic processes $(\eta_t,z_t)$ 
as the unique solution.} The latter fact is important since the noise generating 
process $z_t$ is usually unknown.

\section*{ACKNOWLEDGMENTS}

R.C. acknowledges partial financial support from FAPESP (Grant No. 2013/01242-8).

\vfill

\bibliography{GeneNets_BSDE2}

\end{document}